\renewcommand{\b}[1]{\mbox{\boldmath $#1$}}

\def\cal#1{{\cal #1}}

\def\m@th{\mathsurround=0pt}
\def\n@space{\nulldelimiterspace=0pt \m@th}
\def\biggg#1{{\mbox{$\left#1\vbox to 20.5pt{}\right.\n@space$}}}

\def\beginenum{\begin{enumerate}}
\def\endenum{\end{enumerate}}
\def\bitem{\begin{itemize}}
\def\eitem{\end{itemize}}
\def\bray{\begin{array}}
\def\eray{\end{array}}
\def\begindoc{\begin{document}}
\def\enddoc{\end{document}}
\def\bq{\begin{equation}}
\def\eq{\end{equation}}
\def\bqy{\begin{eqnarray}}
\def\eqy{\end{eqnarray}}
\def\bqyn{\begin{eqnarray*}}
\def\eqyn{\end{eqnarray*}}
\def\bc{\begin{center}}
\def\ec{\end{center}}
\def\bfll{\begin{flushleft}}
\def\efll{\end{flushleft}}
\def\bflr{\begin{flushright}}
\def\eflr{\end{flushright}}
\newcommand{\Avec}{\mbox{\boldmath $A$}}
\newcommand{\Bvec}{\mbox{\boldmath $B$}}
\newcommand{\Evec}{\mbox{\boldmath $E$}}
\newcommand{\Fvec}{\mbox{\boldmath $F$}}
\newcommand{\Gvec}{\mbox{\boldmath $G$}}
\newcommand{\Rvec}{\mbox{\boldmath $R$}}
\newcommand{\Uvec}{\mbox{\boldmath $U$}}
\newcommand{\Vvec}{\mbox{\boldmath $V$}}
\newcommand{\evec}{\mbox{\boldmath $e$}}
\newcommand{\jvec}{\mbox{\boldmath $j$}}
\newcommand{\kvec}{\mbox{\boldmath $k$}}
\newcommand{\nvec}{\mbox{\boldmath $n$}}
\newcommand{\uvec}{\mbox{\boldmath $u$}}
\newcommand{\vvec}{\mbox{\boldmath $v$}}
\newcommand{\wvec}{\mbox{\boldmath $w$}}
\newcommand{\xvec}{\mbox{\boldmath $x$}}
\newcommand{\omegavec}{\mbox{\boldmath $\omega$}}
\newcommand{\Omegavec}{\mbox{\boldmath $\Omega$}}

\documentclass[twocolumn,showpacs,aps]{revtex4}
\usepackage{amsfonts}
\usepackage{amsmath}
\usepackage{amssymb}
\usepackage{color}
\usepackage{graphicx}%
\providecommand{\U}[1]{\protect\rule{.1in}{.1in}}

\begin{document}

\title{Beltrami--Bernoulli Equilibria in Plasmas with Degenerate Electrons}

\author{V. I. Berezhiani}
\email{vazhab@yahoo.com} \affiliation{Andronikashvili Institute of
Physics, TSU, Tbilisi 0177, Georgia }
\affiliation{School of
Physics, Free University of Tbilisi, Georgia}
\author{N. L. Shatashvili}
\email{shatash@ictp.it} \affiliation{Andronikashvili Institute of
Physics, TSU, Tbilisi 0177, Georgia }
\affiliation{Department of
Physics, Faculty of Exact and Natural Sciences, Ivane
Javakhishvili Tbilisi State University (TSU), Tbilisi 0179,
Georgia}
\author{S. M. Mahajan}
\email{mahajan@mail.utexas.edu} \affiliation{Institute for Fusion
Studies, The University of Texas at Austin, Austin,Tx 78712}

\pacs{52.27.Ny, 52.30.Ex, 52.35.We, 96.60.Hv, 97.10.Ex, 97.20.Rp}

\begin{abstract}

A new class of Double Beltrami--Bernoulli equilibria, sustained by
electron degeneracy pressure, are investigated.  It is shown that
due to electron degeneracy, a nontrivial Beltrami--Bernoulli
equilibrium state is possible even for a zero temperature plasma.
These states are, conceptually, studied to show the existence of
new energy transformation pathways  converting, for instance, the
degeneracy energy into fluid kinetic energy. Such states may be of
relevance to compact astrophysical objects like white dwarfs,
neutron stars etc.

\end{abstract}

\startpage{1}
\endpage{1}
\maketitle

\section{Introduction}

Constrained minimization of fluid energy with appropriate helicity
invariants has provided a variety of extremely interesting
equilibrium configurations that have been exploited and found
useful for understanding laboratory as well as astrophysical
plasma systems (see e.g.
\cite{relaxed,woltjer,taylor,sudan,dewar,dewar2} and references
therein). Two particularly simple manifestations of this genre of
equilibria (called Beltrami states) are: 1) The single Beltrami
state, $ \nabla \times {\bf B}=\alpha {\b B}$, discussed by
Woltejr and Taylor \cite{woltjer,taylor} in the context of force
free single fluid magnetohydrodynamics (MHD), and  2) a more
general Double Beltrami State accessible  to Hall MHD -- a
two-fluid system of ions and inertialess electrons \cite{DB}; the
latter has been investigated, in depth, by Mahajan and co-workers
\cite{MY,mmns,osym,sm,mnsy,msms,DBwaves1}. The Beltrami condition
implies an alignment of the fluid vorticity and its velocity, and
the characteristic number of a state is determined by the number
of independent single Beltrami systems needed to construct it.

The Beltrami conditions must be buttressed by an appropriate
Bernoulli constraint to fully describe an  equilibrium state; it
is, then, more descriptive to call them Beltrami-Bernoulli (BB)
states.

Although the BB class of equilibria have been studied for
both relativistic and non-relativistic plasmas, most
investigations are limited to what may be termed  "dilute" or
non-degenerate plasmas so that the constituent particles are
assumed to obey the classical Maxwell-Boltzman statistics. It is
natural to enquire how such states would change/transform if the
plasmas were highly  dense and degenerate (the mean inter-particle
distance is smaller than the de Broglie thermal wavelength) so
that their energy distribution was dictated by Fermi-Dirac
statistics. Notice that at very high densities, the particle Fermi
Energy can become relativistic and the degeneracy pressure may
dominate the thermal pressure.

Such highly dense/degenerate plasmas are found in several
astrophysical and cosmological environments as well as in the
laboratories devoted to inertial confinement and high energy
density physics; in the latter intense lasers are employed to
create such extreme conditions
\cite{Yanovski,tajima1,Dunne,tajima2}. Compact astrophysical
objects like white and brown dwarfs, neutron stars, magnetars with
believed characteristic electron number densities $\sim 10^{26} -
10^{32}$ $cm^{-3}$ , formed under extreme conditions, are the
natural habitats for dense/degenerate matter
\cite{Chandra1,Chandra2,Compact,michel,White,Beloborodov,Shukla}.

\bigskip

In this paper, we develop the simplest model in which the effect
of quantum degeneracy on the nature of the BB class of equilibrium
states can be illustrated. Emphasizing the quantum degeneracy
effects, the aim of this paper is complementary to that of
\cite{MA-spin,BAM-spin} in which the BB like states of a neutral
fluid are investigated when another quantum phenomenon -- the spin
vorticity -- plays a fundamental role. We will choose a model
hypothetical system (later we would show its relevance to specific
aspects of a white dwarf (WD)) of a two-species neutral plasma
with non-degenerate non relativistic ions, and degenerate
relativistic electrons embedded in a magnetic field. To make the
model conform closely to the standard Double BB system, it will be
further assumed that, despite the relativistic mass increase, the
electron fluid vorticity is negligible compared to the electron
cyclotron frequency (such a situation may pertain, for example, in
the pre-WD state of star evolution, and in the dynamics of the WD
atmosphere). The study of the degenerate electron inertia effects
on the Beltrami States in dense neutral plasmas will constitute
the scope of a future publication.

\section{Model}

For an ideal isotropic degenerate Fermi gas of electrons at
temperature $T_e$, the relevant thermodynamic quantities -- the
pressure ${\cal{P}}_e$ and the proper internal energy density
${\cal{E}}_e$ (the corresponding enthalpy, $w_e = {\cal{E}}_e +
{\cal{P}}_e$) \ per unit volume  -- can be calculated to be
\cite{Russo,boltzmann}
\begin{equation}
{\cal{P}}_e=\frac{m_e^{4}c^{5}}{3\pi^{2}\hbar^{3}}f\left(
P_F\right) , \label{B3}
\end{equation}
\begin{equation}
{\cal{E}}_e=\frac{m_e^{4}c^{5}}{3\pi^{2}\hbar^{3}}\left[
P_F^{3}\left( \left( 1+P_F^{2}\right)^{1/2} - 1\right) - f\left(
P_F\right) \right] , \label{B4}
\end{equation}
\noindent where
\begin{equation}
8f\left(P_F\right)=3\sinh^{-1}P_F + P_F\left(
1+P_F^{2}\right)^{1/2}\left( 2P_F^{2}-3\right)   \label{B5}
\end{equation}
and  \ $P_F = p_{F}/m_ec$ \ is the normalized Fermi momentum of
electrons; the Fermi energy may be expressed in terms of \ $P_F$ \
as \ $\epsilon_{F}=m_ec^{2}\left[ \left( 1+P_F^{2}\right )
^{1/2}-1\right]$ . It is useful to note that \ $p_{F}$ \ is
related to the rest-frame electron density \ $n_e$ \ via \
$p_{F}=m_ec\left( n_e/n_{c}\right)^{1/3}$ , where \
$n_{c}=5.9\times10^{29}cm^{-3}$ \ is the critical number-density
at which the Fermi momentum equals \ $m_{e}c$ \ \cite{Akbari}, and
defines the onset of the relativistic regime.  The electron plasma
is treated as the completely degenerate gas -- the thermal energy
of electrons is much lower than their Fermi energy \
($n_eT_e/{\cal{P}}_e\ll 1$) . 
The distribution function of electrons remains locally
Juttner-Fermian which for zero temperature case leads to the just
density dependent thermodynamical quantities \ ${\cal{E}}_e(n_e),
\ {\cal{P}}_e(n_e)$ and \ $w_e(n_e)$. All these quantities
implicitly depend on space-time coordinates via \
$n_e=N_e/\gamma_e $, where \ $N_e$ \ is the density in laboratory
frame of the electron-fluid; \ $\gamma_e=(1-V_e^{2}/c^{2})^{-1/2}$
\ is the Lorentz factor. The electron plasma dynamics is
isentropic and, consequently, obeys the thermodynamical relation
$d(w_e/n_e ) = (d{\cal{P}}_e)/n_e$. Applying this relation, and
after straightforward algebra (see e.g.
\cite{BMYO-Sat,gurovich,BST}), the equation of motion for
degenerate electron fluid reduces to:
\[
\frac{\partial}{\partial t}\left( \sqrt{1+P_F^2}\
\mathbf{p}_e\right) +m_ec^{2}\mathbf{\nabla }\left(
\sqrt{1+P_F^2}\ \gamma_e\right) =
\]
\begin{equation}
= - e\mathbf{E} - \frac{e}{c}\,{\bf V}_e \times {\bf B} +
\frac{e}{c}\, {\bf V}_e \times \nabla \times \left(
\sqrt{1+P_F^2}\ \mathbf{p}_e \right) \ \label{B6}
\end{equation}
with \ $\mathbf{p}_e=\gamma_e m_e\mathbf{V}_e$ \ being electron
hydrodynamical momentum and under our assumption of negligible
electron fluid vorticity the last term can be negligible. For the
non-degenerate ion fluid we have the equation of motion written as
\ ($m_i$ \ is a proton mass):
\[
m_i\left[\frac{\partial {\bf V}_i}{\partial t} + ({\bf V}_i\cdot
\nabla){\bf V}_i\right] = -\frac{1}{N_i}\nabla p_i \ +
\]
\begin{equation}
+ \ e{\bf E} \ + \frac{e}{c}\,{\bf V}_i\times {\bf B} \ .
\label{B6i1}
\end{equation}

\bigskip

Since this short paper is devoted to bringing out the simplest
effects of electron degeneracy on BB states (that may be
very useful in understanding some aspects of the appropriate
astrophysical objects and their evolution), we will borrow
verbatim most of the results for the electron and ion dynamics
\cite{DB,mmns,osym}. For non relativistic ions, and inertialess
electrons, there are two independent Beltrami conditions (aligning
the electron and ion generalized vorticities along their
respective velocities):
\begin{equation}
{\bf b} = a\,N\,\left[{\bf V} - \frac{1}{N}\nabla \times {\bf b}
\right] \ , \label{DB1}
\end{equation}
\begin{equation}
{\bf b}+\nabla \times {\bf V} = d\,N\,{\bf V} \ , \label{DB2}
\end{equation}
where \ ${\bf b}=e{\bf B}/m_ic$ \ and it was assumed, that
electron and proton densities are nearly equal - $N_e\simeq N_i=N$
; here \ $a$ \ and \ $d$ \ are dimensionless constants related to
the two invariants: the magnetic helicity \ $h_1=\int{({\bf
A}\cdot {\bf b})\,d^3x}$ \ and the generalized helicity \
$h_2=\int{({\bf A}+{\bf V})\cdot ({\bf b}+\nabla \times {\bf V})
\,d^3x}$ \ of the system; here \ ${\bf A}$ \ is the dimensionless
vector potential.

Notice that, following the conventional treatments, we have
written our  equations in terms of  normalized one fluid
variables: the fluid velocity \ $\bf V$ \  and the current \ ${\bf
J}=\nabla \times {\bf b}$ \ (via Ampere's law) in terms of which,
the electron and the ion speeds are given by ${\bf V}_e= {\bf
{V}}-(1/N)\nabla\times {\bf b}$, and \ ${\bf V}_i={\bf V}$ ,
respectively (the electrons are assumed to be inertia less). In
this approximation of inertia less electrons, the electron
vorticity is primarily magnetic \ (${\bf b}$) \ while the ion
vorticity has both kinematic and magnetic parts \ (${\bf b}+\nabla
\times {\bf V}$) .

In the preceding equations, the density is normalized to \ $N_0$ \
(the corresponding rest-frame density is \ $n_0$); the magnetic
field is normalized to some ambient measure \ $B_0$; all
velocities are measured in terms of the corresponding Alfv\'en
speed \ $V_A=B_0/\sqrt{4\pi N_0m_i}$ ; all lengths [times]  are
normalized to the skin depth \ $\lambda_i \ [{\lambda_i}/{V_A}]$ ,
where $\lambda_i = c/\omega_{pi}=c\,\sqrt{{m_i}/{4\pi N_0e^2}}$  .

\bigskip

As mentioned in the introduction, the Beltrami conditions
(\ref{DB1}) and (\ref{DB2}) must be supplemented by the Bernoulli
constraint to define an equilibrium state (the stationary solution
of the dynamical system). In the present context, the constraint
reads as
\begin{equation}
\nabla \left(\beta_0\,{\rm{ln}}\,N + \mu_0\,\sqrt{1+P_F^2}\ \gamma
+ \frac{V^2}{2}\right) = 0 \ \label{Bernoulli}
\end{equation}
where $\beta_0$ is the ratio of the thermal pressure to the
magnetic pressure, and $\mu_0=m_ec^2/m_iV_A^2$ \  and for the
electron fluid Lorentz factor we put \ $\gamma_e \simeq
\gamma({\bf V})$ . Stated equivalently, Bernoulli condition
(\ref{Bernoulli}) is an expression of the  balance of all the
remaining potential forces when  Beltrami conditions (\ref{DB1}),
(\ref{DB2}) are imposed on the two-fluid equilibrium equations.

\bigskip

Since \ $P_F = p_F/m_e\,c=(NN_0/n_c\gamma)^{1/3}$ \
[$=(Nn_0/n_c)^{1/3}$] \ is a function of the density \ $N$ , the
system of equations (\ref{DB1})-(\ref{DB2})-(\ref{Bernoulli})
forms a fully specified equilibrium -- a complete system to
determine \ $\bf{N}$ , ${\bf V}$ , and \ $\bf{b}$ . Notice that
the equilibrium continuity equation \ [$\nabla \cdot ({N{\bf
V}})=0$] \ and the divergence free condition for magnetic field \
[$\nabla \cdot {\bf b}=0$] \ are automatically satisfied. The
simplest double BB equilibrium configuration in plasmas with
degenerate electrons has following
noteworthy features: \\

\noindent 1) The Beltrami conditions reflect the simple physics:
(i) the inertia-less (despite the relativistic increase in mass)
degenerate electrons follow the field lines, (ii) while the ions,
due to their finite inertia, follow the magnetic field modified by
the fluid vorticity. The combined field \ ${\bf b}+\nabla \times
{\bf V}$ , an expression of magneto-fluid unification, may be seen
either as an effective
magnetic field or an effective vorticity. \\

\noindent 2) The Beltrami conditions (\ref{DB1}) and (\ref{DB2})
are not directly affected by the degeneracy effects in the current
approximation neglecting the electron inertia. In fact, these are
precisely the two conditions that define the Hall MHD states. In
the highest density regimes, however,  the Fermi momentum (and
hence the Lorentz factor $\gamma ({\bf V})$) may be so large that
the effective electron inertia will have to be included
in (\ref{DB1}), the electron Beltrami condition. \\

\noindent 3) In this minimal model, the electron degeneracy
manifests, explicitly, only through the Bernoulli condition
(\ref{Bernoulli}). The degeneracy induced term, proportional to \
$\mu_0$ \, would go to unity (whose gradient is zero), and would
disappear in the absence of the  degeneracy pressure. For
significant \ $P_F$ , on the other hand, the degeneracy pressure
can be far bigger than the thermal pressure (measured by \
$\beta_0$) . In fact, the degenerate electron gas, can sustain a
qualitatively new state: a nontrivial Double Beltrami--Bernoulli
equilibrium at zero temperature. In the classical zero-beta
plasmas, only the relatively trivial, single Beltrami states are accessible \cite{ymois}. \\

\noindent 4) It is trivial to eliminate \ ${\bf b}$ \ in Eqs.
(\ref{DB1}) and (\ref{DB2}) to obtain
\begin{equation}
\frac{1}{N}\nabla \times \nabla \times {\bf V} + \nabla \times
\left[ \left( \frac{1}{aN}-d \right)N\,{\bf V} \right] + \left(
1-\frac{d}{a} \right){\bf V} = 0 \ , \label{DB3}
\end{equation}
which, coupled with (\ref{Bernoulli}), provides us with a closed
system of four equations in four variables \ $(N, \, {\bf V})$ .
Once this is solved with appropriate boundary conditions, one can
invoke (\ref{DB2}) to calculate ${\bf b}$ . The reader can find
the solution for the similar mathematical problem
relevant to the non-degenerate case (in the context of solar atmosphere) in \cite{mmns}. \\

\noindent 5) The Bernoulli  condition (\ref{Bernoulli}) introduces
a brand new player in the equilibrium balance; the spatial
variation in the electron degeneracy energy (proportional to
$\mu_0$) could increase or decrease the plasma $\beta_0$ or the
fluid kinetic energy (measured by $V^2$) in the corresponding
region. Thus Fermi energy could be converted to kinetic energy; it
could also forge a re-adjustment of the kinetic energy from a
high-density/low-velocity plasma to a low-density/high-velocity
plasma. Similar energy transformations, mediated through classical
gravity, were discussed in Mahajan {\it et al} (2002; 2006).

\bigskip

The extensions as well as a detailed analysis of
(\ref{Bernoulli}-\ref{DB3}) are under investigation. For instance,
when electron fluid degeneracy is very high and one can not
neglect inertia effects in their generalized vorticity, the order
of BB states is likely to rise; such higher order states (like the
triple BB state when electron inertia is retained) have been
studied for specific cases \cite{YM-2,SM-TB,iqbal,pino}. Another
natural extension for  the current formalism (supper-relativistic
electrons) will be the introduction of Gravity, which, in
principle, could  balance the highly degenerate electron fluid
pressure.  Gravity (Newtonian) effects in the BB system have been
investigated in the solar physics context (e.g Mahajan {\it et.al}
(2002), Mahajan {\it et. al.} (2005,2006); for disk-jet structure
formation -- \cite{disk-jet1,disk-jet2}).

Since our aim in this paper is  merely demonstrating the
possibility of Beltrami--Bernoulli equilibria sustained by
electron degeneracy pressure, we will not work out the detailed
solutions of Eqs. (\ref{Bernoulli}-\ref{DB3}). Because this
system, in its non-degenerate form, has been highly studied
(\cite{DB,mmns,osym,DBwaves1,DBwaves2} and references therein), we
can safely draw interesting inferences about:

1) Some distinguishing features of the expected
"degeneracy-modified" solutions and even the significance and
possible applications of somewhat straightforward extensions of
these solutions (keeping electron inertia and adding gravity, for
example). Several of these general features have already
discussed.

2) possible physical systems where such equilibrium solutions
may find relevance.

A possible application of the "degenerate" BB states may be found
in stellar physics. Here is a short summary of the relevant
phenomenology:

It is well-known that when a star collapses, and cools down, the
density of lighter elements increases affecting the total
pressure/enthalpy of unit fluid element -- first order departure
from the classical e-i plasma; beyond the hot, pre-white dwarf
stage, photon cooling dominates and gravitational contraction is
dramatically reduced as the interior equation of state hardens
into that of a strongly degenerate electron gas.
The degenerate electrons provide the dominant pressure, while the 
contribution of thermal motion of ions into the pressure is negligible 
(see the review \cite{WDReview} and references therein).

Recent studies show that a significant fraction of White Dwarfs
are found to be magnetic with typical fields strengths below 1KG.
Massive and cool White Dwarfs, interestingly, are found with much
higher fields detected (see \cite{WDmagnetism} and references
therein). It is argued that the origin of magnetic fields in WD
stars may be linked to possible field-generating merger events
preceding the birth of the white dwarf. On the other hand, Wegg \&
Phinney (2012) concluded that the kinematics of massive WDs are
consistent with the majority being formed from single star
evolution. In \cite{WDmagnetism} it was shown that WD stars with
such surface temperatures that convection zones develop, seems to
show stronger magnetic fields than hotter stars;  the mean mass of
magnetic stars seems to be on average larger than the mean mass of
non-magnetic WD stars. Recent investigations (\cite{old-WD-B} and
references therein) have uncovered several cool, magnetic,
polluted hydrogen atmosphere (DAs) white dwarfs. It was found that
the incidence of magnetism in old, polluted white dwarfs (DAZ)
significantly exceeds what is found in the general white dwarf
population suggesting a hypothetical link between a crowded
planetary system and magnetic field generation. Polluted white
dwarfs provide an opportunity to investigate the ultimate fate of
planetary systems and, hence, it is of crucial importance to study
the origin and evolution of surface magnetic fields of such
DAZ-es.

\bigskip

Let us now explore, through a simple example, if  degenerate BB
states could shed some light on the physics of WDs. Considering
High magnetic field white dwarfs, we assume: the degenerate
electrons densities $\sim (10^{25}-10^{29})\,cm^{-3} $ \ ;
magnetic fields $\sim (10^5-10^{9})\,G $ \ , and temperatures \
$\sim (40000-6000)\,K $ \ . For these  parameters, the Alfv\'en
speed\ $V_A\sim 
(10^4-10^6)\,cm/s$ \ , yielding \
 $\beta_0 \sim (10^6-10^0)$ \ and \ $\mu_0 \sim (10^{10}-10^6)\gg 1$.
The ion skin-depth \ $\lambda_i \sim 
(10^{-5}-10^{-7})\,cm  $ \ turns out to be rather short.

For this class of systems, the second term (degeneracy pressure)
in Eq.(\ref{Bernoulli}) is always much larger than the first term,
the thermal pressure. Neglecting the first term, and remembering
that for  non relativistic flows (essential at ion speeds)\
$\gamma ({\bf V})\sim 1$ \ , Eq.(\ref{Bernoulli}) -- Bernoulli
Condition -- with inclusion of classical (Newtonian) gravity
(justified by observations for WDs) implies
\begin{equation}
\mu_0\,\sqrt{1+P_F^2} - \frac{R_A}{R} + \frac{V^2}{2} = const
\label{WDbalance}
\end{equation}
where the $const $ measures, in some sense,  the main energy
content of the fluid; the Beltrami conditions (\ref{DB1}),
(\ref{DB2}) remain the same; \ $R$ \ is a radial distance from the
center of WD normalized to its radius \ $R_{W}\ [\sim (0.008 -
0.02)\ R_{\odot}]$ \ and $R_A=GM_{W}/R_{W}V_A^2$ (here $G$ is the
gravitational constant and $M_W$ - WD mass). Since \ $P_F$ \ is a
function of Fermi energy (and hence, of density), we assume that
at some distance \ $R_*$ \
(corresponding to density maximum), \ $P_F$ \ reaches its maximum
value \ $P_{F*}$. Taking the corresponding
minimum velocity to be zero \ ($V_{*}\sim 0$),  we find \ $const
=\mu_0\,\sqrt{1+P_{F*}^2} - R_A/R_*$. The magnitude of the
velocity is now determined to be
\begin{equation}
|{\bf V}|\sim
\sqrt{2\mu_0}\ \kappa (P_F) \ \label{V-estim}
\end{equation}
with
\begin{equation}
\kappa (P_F)=\left[\,\left(\sqrt {1+P_{F*}^2 } - \sqrt{1+P_F^2}\
\right) - \frac{R_A}{\mu_0}\left(\frac{1}{R_*}-\frac{1}{R}
\right)\,\right]^{1/2} \ . \label{kappa}
\end{equation}

\vspace{0.5cm}

\noindent Notice that the dimensionless coefficient \ $R_A/ \mu_0 \ll
1$ \ measures the relative strength of  gravity versus the
degenerate pressure term. For WD-s with Mass \ $M_{W}\sim (0.8 -
0.25)\,M_{\odot}$ \ and radius \ $R_{W}\sim (0.013 -
0.02)\,R_{\odot}$ , \ $R_A/\mu_0 \sim (0.2 - 0.04) \ll 1$; \ less
massive the WD, the  smaller is the coefficient.

In addition, the DB structure scales are small compared to
$R_W$ in outer layers of the WD (where our model applies). The
gravity contribution to the flow velocity (at specific distance of
outer layers of WD-s with $R\geq R_*$ and $(R-R_*)/R_*\ll 1 \
, R_*\leq 1$ ), therefore, can be readily neglected. The gravity
contribution, exactly like in the solar case \cite{mnsy,msms},
determines the radial distance in WD's outer layer over which the
"catastrophic" (fast) acceleration of flow may appear (due
to the magneto-fluid coupling). In the regions where the flows are
insignificant (at very short distances from the WD's surface)
gravity controls the stratification but as we approach the 
flow "blow-up" distances (the flow becomes strong) the
self-consistent magneto-Bernoulli processes take over and control
the density (and hence the velocity) stratification.

Calculating the maximum flow velocity, occurring at $\kappa(P_F)$
maximum (density  minimum), needs  a detailed knowledge of the
system. One does, however,  notice  that if \
$\sqrt{2\mu_0}\,\kappa(P_F) > 1$ \, the generated flow is locally
super-Alfv\'enic in contradistinction to the non-degenerate,
thermal pressure dominated plasma, when the maximal velocity due
to the magneto-Bernoulli mechanism be locally sub-Alfv\'enic when
local plasma beta $<1$ as in the Solar Atmosphere. This simple
example shows that the electron degeneracy effects can be both
strong, and lead to interesting predictions like the
anticorelation between the density and flow speeds.

The richness introduced by electron-degeneracy to the the
Beltrami-Bernoulli states could help us better understand compact
astrophysical objects. When the star contracts, for example, its
outer layers keep the multi-structure character although density
in the structures becomes defined by electron degeneracy pressure.
Then, important conclusion for future studies is that when
studying the evolution of the atmospheres/outer layers of compact
objects, flow effects can not be ignored. More specifically, the
knowledge of the effects introduced by flows (observed in stellar
outer layers) acquired for classical plasmas can be used when
investigating the dynamics of White Dwarfs and their evolution.

\section{Conclusions}

In conclusion, in the present paper we found the possibility of
the existence of Double Beltrami relaxed states in plasmas with
degenerate electrons (often met in astrophysical conditions).
Since non degenerate double BB states guarantee scale separation
phenomenon that, for appropriate conditions, provide energy
transformation pathways for various astrophysical phenomena
(erruptions, fast/transient outflow and jet formation, magnetic
field generation, structure formation, heating and etc.), such
pathways could be easily explored for the degenerate case with
degeneracy pressure providing an additional energy source.
Particularly interesting could be finding the fate of a Star, when
contracting and cooling, and becoming a White Dwarfs since the
latter is assumed to be a boundary condition for Stellar
Evolution. Our future studies will be devoted to detailed
investigation of present model to explain the existence of
large-scale structures (like surface magnetic fields, flows and
outflows, erruptions) in astrophysical objects with degenerate
plasmas as well as to explore the evolution of multi-structure
stellar outer layers when contracting, cooling.

\bigskip

Authors express their thanks to N.L. Tsintsadze for valuable
discussions. Authors acknowledge special debt to the Abdus Salam
International Centre for Theoretical Physics, Trieste, Italy. The
work of S.M.M. was supported by USDOE Contract No. DE–FG
03-96ER-54366.

\end{document}